\documentclass{aastex}
\usepackage{emulateapj5}
\usepackage{graphics}
\usepackage{psfig}

\def\R{{\sl ROSAT}}
\def\A{{\sl ASCA}}

\def\B{{\sl BeppoSAX}}
\def\X{{\sl XMM-Newton}}
\def\H{{\sl HST}}

\def\gs{\mathrel{\mathchoice {\vcenter{\offinterlineskip\halign{\hfil
$\displaystyle##$\hfil\cr>\cr\sim\cr}}}
{\vcenter{\offinterlineskip\halign{\hfil$\textstyle##$\hfil\cr
>\cr\sim\cr}}}
{\vcenter{\offinterlineskip\halign{\hfil$\scriptstyle##$\hfil\cr
>\cr\sim\cr}}}
{\vcenter{\offinterlineskip\halign{\hfil$\scriptscriptstyle##$\hfil\cr
>\cr\sim\cr}}}}}
\def\ls{\mathrel{\mathchoice {\vcenter{\offinterlineskip\halign{\hfil
$\displaystyle##$\hfil\cr<\cr\sim\cr}}}
{\vcenter{\offinterlineskip\halign{\hfil$\textstyle##$\hfil\cr
<\cr\sim\cr}}}
{\vcenter{\offinterlineskip\halign{\hfil$\scriptstyle##$\hfil\cr
<\cr\sim\cr}}}
{\vcenter{\offinterlineskip\halign{\hfil$\scriptscriptstyle##$\hfil\cr
<\cr\sim\cr}}}}}

\begin{document}

\title{Probing the Complex and Variable X-ray Absorption of Markarian~6 with {\sl XMM-Newton}}

\author{Stefan Immler,\altaffilmark{1} W. N. Brandt,\altaffilmark{1} 
Cristian Vignali,\altaffilmark{1} Franz E. Bauer,\altaffilmark{1} \\
D. Michael Crenshaw,\altaffilmark{2} John J. Feldmeier,\altaffilmark{3}
and Steven B. Kraemer\altaffilmark{4}}

\altaffiltext{1}{Department of Astronomy \& Astrophysics, 525 Davey Laboratory, 
	The Pennsylvania State University, University Park, PA 16802}
\altaffiltext{2}{Department of Physics and Astronomy, Georgia State University, Atlanta, GA 30303}
\altaffiltext{3}{Department of Astronomy, Case Western Reserve University, 
	10900 Euclid Avenue, Cleveland, OH 44106}
\altaffiltext{4}{Catholic University of America and Laboratory for Astronomy and Solar
  Physics, NASA's Goddard Space Flight Center, Code 681, Greenbelt, MD, 20771}

\shorttitle{Probing the Complex and Variable X-ray Absorption of Mrk~6 with {\sl XMM-Newton}}
\shortauthors{Immler et al.}
\begin{abstract}
                                            
We report on an X-ray observation of the Seyfert~1.5 galaxy Mrk~6 
obtained with the EPIC instruments onboard \X. Archival \B\ PDS data 
from \hbox{18--120~keV} were also used to constrain the underlying 
hard power-law continuum. The results from our spectral analyses 
generally favor a double partial-covering model, although other spectral 
models such as absorption by a mixture of partially ionized and neutral gas 
cannot be firmly ruled out. Our best-fitting model consists of a power law 
with a photon index of $\Gamma=1.81^{+0.22}_{-0.20}$ and partial covering with 
large column densities up to $N_{\rm H}\sim10^{23}~{\rm cm}^{-2}$. 
We also detect a narrow emission line consistent with Fe K$\alpha$ fluorescence 
at $6.45^{+0.03}_{\rm -0.04}$~keV with an equivalent width of 
$93^{+26}_{\rm -20}$~eV. Joint analyses of \X, \A, and \B\ data 
further provide evidence for both 
spectral variability (a factor of $\sim2$ change in absorbing column) and 
absorption-corrected flux variations (by $\sim60$\%) 
during the $\sim4$ year period probed by the observations.

\end{abstract}


\section{Introduction}
\label{introduction}
Markarian 6 (Mrk~6, IC~0450; $z=0.0188$; 77~Mpc distant for 
$H_0=75~{\rm km~s}^{-1}~{\rm Mpc}^{-1}$) is one of a handful of
intermediate Seyferts of type~1.5, including NGC~4151 and Mrk~766, 
showing evidence for an `ionization cone' rather than the `ionization halo'
expected for a type~1 viewing geometry (Meaburn, Whitehead, \& Pedlar 1989;
Kukula et~al.\ 1996). A torus `atmosphere' along the line of sight has 
been suggested to explain the presence of the ionization cone (e.g.,
Evans et al.\ 1993). 
The torus atmosphere is thought to be optically 
thick for ionizing radiation between the Lyman edge and soft X-rays,
but it allows radiation outside of this spectral range (e.g., broad optical
line emission from the Broad Line Region) to pass through with only moderate 
obscuration. Mrk~6 shows significant optical line profile 
variations on time-scales of months to years, suggesting that at 
least some of the gaseous material creating the lines can undergo coherent 
variations (e.g., Rosenblatt et~al.\ 1992; Eracleous \& Halpern 1993).

X-ray investigations of Seyfert galaxies are well suited to probing matter 
along the line of sight toward their central black hole regions. 
Prominent examples of intermediate Seyferts that have been well
studied in X-rays include the `cousins' of Mrk~6, NGC~4151 
(e.g., Weaver et~al.\ 1994; Ogle et~al.\ 2000; Yang, Wilson, \& Ferruit 2001; 
Schurch \& Warwick 2002) and Mrk~766 (e.g., Leighly et~al.\ 1996; Page et~al.\ 
1999, 2001; Matt et~al.\ 2000; Branduardi-Raymont et~al.\ 2001).
In contrast, the only published X-ray study of Mrk~6 is that of 
Feldmeier et~al. (1999; hereafter F99) using 0.6--9.5~keV \A\ data.
These data revealed heavy and complex X-ray absorption 
that was best fit by a double partial-covering model with large column 
densities [$\sim(3$--$20)\times10^{22}~{\rm cm}^{-2}$], likely due to
the torus atmosphere. However, detailed X-ray spectral modeling was
limited by modest photon statistics and the fact that the absorption in
Mrk~6 dominates the X-ray spectrum throughout most of the \A\ bandpass.

We capitalized on the superior throughput (about an order of magnitude higher 
than \A) and good spectral resolution of the instruments onboard \X\ 
to investigate further the X-ray emission and absorption properties of 
Mrk~6. We also used the \X\ data to perform sensitive searches for temporal 
and spectral variability during the observation and in comparison with 
earlier \A\ and \B\ observations.

\section{X-ray Observations and Data Reduction}
\label{obs}

\subsection{\X}

Mrk~6 was observed with the European Photon Imaging Camera (EPIC)
onboard \X\ (Jansen et~al.\ 2001) with on-source exposure times of 
31.8~ks (EPIC p-n) and 30.8~ks (EPIC MOS~1 and MOS~2). 
It was also observed with the Reflection Grating Spectrometer (RGS) for 46.4~ks 
(RGS~1) and 37.9~ks (RGS~2). The EPIC data allow sensitive \hbox{0.2--12~keV}
imaging spectroscopy with moderate spectral ($E / \Delta E \sim20$--$50$) 
and angular ($6''$ FWHM) resolution. The parameters of the individual 
observations are listed in Tab.~\ref{obs_tab}.

Both the p-n and MOS data were acquired in `full-frame mode'
using the medium filter.
The data were reduced using the Science Analysis System (SAS) v5.3.3
with the latest calibration products.
Due to background flares at the end of the observation period, 7\% of 
the p-n and 31\% of the MOS~1 and MOS~2 data were rejected. 
After inspection of a source-free background spectrum extracted 
from CCD 1 on the p-n and MOS detectors, we selected data only in the 
0.3--12~keV energy range due to enhanced background
at low photon energies (up to a factor of $\sim5$).
We furthermore excluded bad and hot pixels from the data. 
Since the cleaned RGS data did not yield enough photon statistics 
for spectral analysis, they were not used further.

Source counts were extracted for the p-n and MOS detectors within
$30''$-radius circular regions (the 85\% encircled-energy radius at 7.5~keV).
The background was extracted locally from source-free regions. For the p-n we 
used a rectangular region of size $2\farcm4\times3\farcm8$ close to Mrk~6 on 
the same p-n chip, and for the MOS we used an annulus with inner and outer 
radii of $2\farcm5$ and $4\farcm3$. Given the observed count rates of 
$\sim 1.2$ and $\sim 0.4~{\rm cts~s}^{-1}$ for the p-n and MOS detectors,
the pile-up fractions and dead times are estimated to be $\ls 2$\%.
The source counts were binned with a minimum 
of $50$ counts per bin to allow $\chi^2$ spectral fitting. 
We also constructed background-corrected light curves 
of Mrk~6 in the soft (0.3--2~keV), hard (2--12~keV), and broad 
(0.3--12~keV) bands by binning the events into 250~s intervals
and excluding telemetry dropouts.

\subsection{\B}

Archival \B\ MECS 2+3 (Medium Energy Concentrator Spectrometer) 
and LECS (Low Energy Concentrator Spectrometer) data are also
used in the analyses below (see Tab.~\ref{obs_tab}); these data 
cover the 1.3--10~keV (MECS) and 0.5--4.5~keV (LECS) energy bands.
To extend the spectral range of our analyses to higher
photon energies, which is useful to constrain the underlying
power-law continuum, we furthermore used the \B\ PDS 
(Phoswich Detection System) data in the energy range 18--120~keV.

We checked for contamination of the PDS data due to other X-ray 
sources within its field of view ($1\fdg3$ FWHM).
No strong X-ray source other than Mrk~6 is found within the 
$\sim30'$-diameter EPIC field of view in high-energy ($>5$~keV) 
images. The \R\ All-Sky Survey broad-band source catalog 
shows only two other X-ray sources (NGC~2256 and NGC~2258) located within 
the PDS field of view at large off-axis angles ($23\farcm2$ and $18\farcm1$).
Both of these sources have 2--10~keV MECS fluxes much lower (only $\sim2$--$3\%$) 
than that of Mrk~6. Since the PDS effective area decreases sharply with increasing 
off-axis angle, their contamination to the Mrk~6 PDS spectrum is negligible.

\section{Results}
\label{results}

\subsection{Spectral Analysis}
\label{spec_analysis}

Spectral analysis was performed using XSPEC v11.2 (Arnaud 1996).
All errors are quoted at the $90\%$ level of confidence for one parameter 
of interest ($\Delta \chi^2 = 2.71$; Avni 1976)
and Galactic absorption ($N_{\rm H}=6.4\times10^{20}~{\rm cm}^{-2}$; 
Stark et al.\ 1992) is always implicitly included. 
The previous \A\ observation has shown that Mrk~6 is an intrinsically 
absorbed system ($\gs10^{22}$--$10^{23}~{\rm cm}^{-2}$; F99).
We therefore used the 18--120~keV \B\ PDS data, which should be
much less affected by absorption, to determine the shape of 
the underlying X-ray continuum. Fitting a power-law
model to the PDS data (model~1 in Tab.~\ref{spec_tab}) gives a best-fit 
photon index of $\Gamma=1.81^{+0.22}_{-0.20}$ and an unabsorbed 
20--100~keV flux of $5.0\times10^{-11}~{\rm ergs~cm}^{-2}~{\rm s}^{-1}$. 

We then fitted the \X\ and \B\ PDS spectra simultaneously. 
Due to uncertainties in the cross-calibration of the instruments 
(p-n and MOS: $\ls7\%$; Snowden 2002) as well as possible
variability between the \X\ and \B\ observations, 
we left all normalizations for the spectral components free. 
All other spectral parameters were tied together.
We first tried fitting a power-law model with simple (fully 
covering) intrinsic absorption. Such absorption could arise in 
the host galaxy, perhaps in gas associated with the irregularly 
distributed dust seen in the \H\ image of Malkan, Gorjian, \& Tam (1998). 
It could also arise on smaller scales in the nuclear region. This model 
is statistically rejected ($\chi^2_{\nu}=3.24$; ${\rm d.o.f.=867}$). 
It also gives a photon index ($\Gamma\sim0.9$) well below 
that derived from our \B\ PDS analysis and expected intrinsically 
for a Seyfert galaxy ($\Gamma\sim1.6$--$2.2$; 
e.g., Nandra et~al.\ 1997; Risaliti 2002).

We next tried fitting a power-law model with a single partial-covering 
absorption component. Partially covering absorption could arise from 
electron scattering of X-rays in the nucleus of Mrk~6 (see F99), and such 
absorption has been used successfully to explain the X-ray spectra of other 
intermediate Seyferts. This model also includes the simple intrinsic 
absorption used in the previous model, since again there could be 
absorption in the host galaxy after the X-rays have escaped the
nucleus (this simple intrinsic absorption will be implicitly 
included in all of the following spectral fits). Finally, we 
added a Gaussian emission line to model Fe~K$\alpha$ emission 
visible in the spectrum. This model is marginally acceptable 
($\chi^2_{\nu}=1.08$; ${\rm d.o.f.=859}$) although it leaves
clear systematic residuals below $\sim2$~keV and above $\sim40$~keV. 
Furthermore, like the previous model, it requires an implausibly 
low photon index of $\Gamma\sim1.3$. If we constrain the photon 
index to lie in the range 1.6--2.2, the fit is statistically 
unacceptable. 

Given the failures of the two previous spectral models, we tried a 
model consisting of a power law and two partially covering 
absorbers (model~2 in Tab.~\ref{spec_tab}). This model 
was successfully used by F99 to describe 
the \A\ data, and double partial covering could arise if electron scattering 
of X-rays provides several lines of sight into the nucleus. The photon index 
was fixed to the best-fit value derived from the \B\ PDS data ($\Gamma = 1.81$;
our results are not changed materially if the photon index is left as a free
parameter), and we also included a Gaussian Fe~K$\alpha$ emission line. 
Despite some small systematic residuals below $\sim 2$~keV, a good
overall fit to the \X\ and \B\ data is found ($\chi^2_{\nu}=0.96$; 
${\rm d.o.f.=858}$; see Fig.~\ref{spectrum}). 
The absorption includes a high column density 
($10.96^{+0.58}_{-0.42}\times10^{22}~{\rm cm}^{-2}$) component
covering $(57\pm1)\%$ of the emitting region, as well as 
a lower column density ($2.46^{+0.07}_{-0.05}\times10^{22}~{\rm cm}^{-2}$) 
component with a higher covering percentage of ($93\pm1)\%$.
The best-fit rest-frame energy of the Fe~K$\alpha$ line is 
$6.45^{+0.03}_{\rm -0.04}$~keV with an equivalent width of
$93^{+26}_{\rm -20}$~eV. There is no clear evidence for a 
broad Fe~K$\alpha$ line, and the fitted line properties are 
consistent with those from \A. Given its properties, the
observed line could originate via reprocessing in the 
outer part of the accretion disk (e.g., George \& Fabian 1991) 
or perhaps in Broad Line Region clouds 
(see equation~5 in Eracleous, Halpern, \& Livio 1996).

The Fe~K$\alpha$ line is fairly weak for a Seyfert galaxy, suggesting
that Compton reflection is unlikely to make a dominant contribution 
to the X-ray spectrum. To examine any effects of Compton reflection
further, we replaced the simple power law in the previous fit 
with a power law plus neutral reflection (the `PEXRAV' model 
in XSPEC assuming solar abundances). All other aspects of the model
were unchanged. We obtain a relative reflection fraction 
(defined as $r=\Omega/2\pi$, where $\Omega$ is the angle subtended by 
the reflector) of $r=1.21^{+0.14}_{-0.16}$, 
an inclination angle of cos~$i=0.94^{+0.01}_{-0.25}$, and 
a cut-off energy of $\sim 238$~keV (model~3 in Tab.~\ref{spec_tab}).
The fit quality is slightly improved compared to the previous fit
($\chi^2_{\nu}=0.95$; ${\rm d.o.f.=855}$), but there is no material
change in the nature of the partially covering absorption. 
Due to both statistical and systematic uncertainties associated
with this model, we do not consider the derived inclination angle
to have clear physical significance.

We also tested a model which assumes that the nuclear X-ray
emission is attenuated by both partially ionized
and neutral material, dispersed along the line of sight.
This model was proposed by Schurch \& Warwick (2002) to characterize 
the complex X-ray absorption of NGC~4151, a Seyfert with many 
similarities to Mrk~6 (see \S1). 
Although the model gives a relatively good fit to the 2--120~keV
data, large residuals are visible below $\sim2$~keV
($\chi^2_{\nu}=1.14$; ${\rm d.o.f.=866}$; see Fig.~\ref{spectrum}). 
We obtain an ionization parameter of 
${\rm log}~\xi = {\rm log}~(L_{\rm ion}/nr^2) = 2.42\pm0.01$
(with $L_{\rm ion}$ the source ionizing luminosity in the 0.0136--13.6~keV
band, $n$ the number 
of hydrogen atoms/ions per $\rm cm^{-3}$, and $r$ the distance from
the central source to the photoionized region), an ionized column density 
of $1.33^{+0.01}_{-0.02}\times10^{23}~{\rm cm}^{-2}$, and a neutral
column density of $3.35^{+0.17}_{-0.11}\times10^{21}~{\rm cm}^{-2}$. 

A slightly better fit ($\chi^2_{\nu}=1.09$; ${\rm d.o.f.=863}$) 
is obtained if the simple power law in the previous fit is replaced 
with a power law plus neutral reflection (model~4 in Tab.~\ref{spec_tab}).
As for the double partial-covering model, the reflection is not a 
dominant spectral component and the best-fit absorption parameters
do not change materially. 

We performed several additional fits to check the general 
robustness of our results. Models with no intrinsic absorption 
and Compton reflection cannot provide acceptable fits. Models with 
fully covering intrinsic absorption and Compton reflection can at
best provide marginally acceptable fits. These models also suffer
from physical consistency problems since the energy region near 
the Fe~K$\alpha$ line is dominated by the reflection component. 
An Fe~K$\alpha$ line with a large equivalent width of $\sim1$--$2$~keV 
is then expected (e.g., Matt, Brandt, \& Fabian 1996) but not observed. 

Using the double partial-covering model (model~2 in Tab.~\ref{spec_tab}), 
we derive observed 0.3--2~keV and 2--10~keV \X\ fluxes of 
$5.3\times10^{-13}~{\rm ergs~cm}^{-2}~{\rm s}^{-1}$ and 
$1.2\times10^{-11}~{\rm ergs~cm}^{-2}~{\rm s}^{-1}$, respectively.
These values are consistent with those in \S2.3.5 of F99, 
given statistical and systematic uncertainties. 
The absorption-corrected 2--10~keV flux and luminosity are 
$2.0\times10^{-11}~{\rm ergs~cm}^{-2}~{\rm s}^{-1}$ and 
$1.4\times10^{43}~{\rm ergs~s}^{-1}$, respectively. 
The absorption-corrected 0.3--2~keV flux and luminosity are 
less certain due to the large required absorption correction; 
likely values are 
$\sim 1.5\times10^{-11}~{\rm ergs~cm}^{-2}~{\rm s}^{-1}$ and 
$\sim 1.1\times10^{43}~{\rm ergs~s}^{-1}$, respectively. 
Using the partially ionized and neutral absorption model (model~4 in 
Tab.~\ref{spec_tab}), we derive an absorption-corrected 2--10~keV flux
and luminosity $\sim36\%$ lower than from the double partial-covering model.

\subsection{Temporal and Spectral Variability}
\label{timing_analysis}

After fitting constant models to light curves and inspecting the
residuals, we do not find any rapid, large-amplitude X-ray variability 
during the \X\ or \B\ observations. While constant-model fits 
are statistically inconsistent with the \X\ p-n, \B\ MECS, 
and \B\ LECS light curves, there are no strong, systematic
variations about these fits (the poor fit quality appears to be 
due to stochastic small-amplitude variability only slightly larger 
than the statistical noise). We can constrain the amplitude of any 
systematic variability to be 
$\ls15\%$ for the \X\ p-n,
$\ls17\%$ for the \B\ MECS, and
$\ls29\%$ for the \B\ LECS.
Within the individual observations, no evidence for spectral variability 
is found upon analysis of hardness ratios (e.g., 2--12~keV to 
0.3--2~keV) computed as a function of time. No variability was seen 
in the \A\ data (F99), consistent with our results here. 

\begin{figure}[h!]
\centerline{ {\hfil\hfil
\psfig{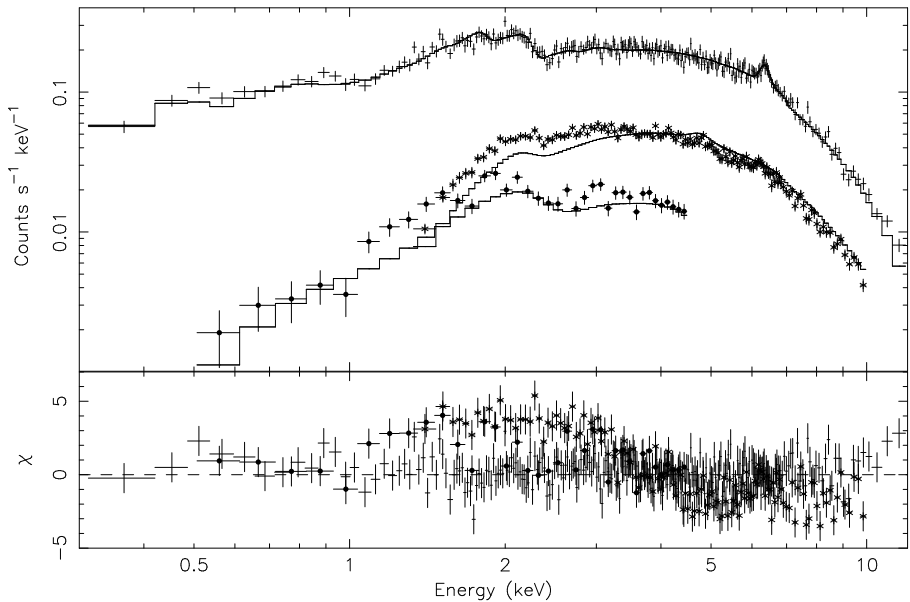}
\hfil\hfil} }
\caption{\X\ EPIC p-n (top; 0.3--12~keV), \B\ MECS (middle; 1.3--10~keV) and LECS 
(lower; 0.5--4.5~keV) spectra of Mrk~6. Counts were binned into energy channels with 
a signal-to-noise ratio of ${\rm S/N}>10$ for plotting purposes. The best-fit double 
partial-covering model is plotted as the histograms (see model~2 in Tab.~\ref{spec_tab}).
The fit residuals are shown in the lower panel (in units of $\sigma$). 
Note the inability to fit the \X\ and \B\ data simultaneously with the double partial
covering model.
\label{spectrum_sax}}
\end{figure}

We have checked for inter-observation spectral variations via joint 
spectral fitting. 
A joint spectral analysis of the \X\ and \A\ data gives a fit with 
$\chi^2_{\nu}=1.04$ (${\rm d.o.f.=1595}$).\footnote{The \A\ spectra used in
this analysis were the same as those presented in F99 (see Tab.~\ref{obs_tab}).} 
This joint fit is statistically acceptable, indicating that we cannot 
prove spectral variability between the \A\ and \X\ observations. 
If we use the \X\ and \B\ data, however, we obtain $\chi^2_{\nu}=1.20$ 
(${\rm d.o.f.=1410}$). This joint fit can be rejected with $>99.9$\% 
confidence, indicating spectral variability. Simultaneous \X\ and \B\
observations of the Seyfert~1 galaxies NGC~5548 (Pounds et al.\ 2003)
and IC~4329A (Gondoin et al.\ 2001) have shown that excellent fits can be 
found in joint spectral analyses, ruling out major cross-calibration 
errors. The spectral differences between the \X\ 
and \B\ data of Mrk~6 are largest below $\sim 3$~keV where
absorption dominates the spectral shape (see Fig.~\ref{spectrum_sax}), so
absorption variability seems a likely explanation. 
We therefore analyzed the \B\ data (MECS, LECS and PDS) individually, 
which gives best-fit column densities a factor of $\sim 2$ smaller compared to 
\X\ ($N_{\rm H}^1 =4.76^{+0.27}_{-0.25} \times 10^{22}~{\rm cm}^{-2}$; 
$N_{\rm H}^2 =1.37^{+0.08}_{-0.07} \times 10^{22}~{\rm cm}^{-2}$)
and covering fractions which are consistent with our \X\ results
[$f_{\rm c}^1 =(54\pm1)\%$; $f_{\rm c}^2 =(95\pm1)\%$].
The observed \hbox{0.3--2~keV} and \hbox{2--10~keV} \B\ fluxes are  
$1.1\times10^{-12}~{\rm ergs~cm}^{-2}~{\rm s}^{-1}$ and 
$2.4\times10^{-11}~{\rm ergs~cm}^{-2}~{\rm s}^{-1}$, respectively.
These values are significantly higher than during the \X\ observation. 
The absorption-corrected 2--10~keV flux and luminosity are 
$3.2\times10^{-11}~{\rm ergs~cm}^{-2}~{\rm s}^{-1}$ and 
$2.1\times10^{43}~{\rm ergs~s}^{-1}$, respectively. 
The 2--10~keV luminosity of Mrk~6 during the \B\ observation was 
$\sim 60$\% higher than during the \X\ observation. 

\section{Discussion and Summary}
\label{discussion}

The analyses above have substantially tightened the constraints upon
the X-ray absorption and emission properties of Mrk~6. The \X\ spectrum 
has $\sim 7$ times as many counts and a wider bandpass than the 
earlier \A\ spectrum, and the 18--120~keV \B\ PDS data have provided
the first reliable determination of the underlying X-ray continuum shape
(a critical quantity for modeling of the X-ray absorption). The 
absorption measured in the new data can be fit acceptably with 
the same basic double partial-covering model used to fit the \A\ 
data (F99), providing substantially improved support for the applicability
of this model to Mrk~6. The measured column densities are large with
$N_{\rm H}$ up to $\sim 10^{23}$~cm$^{-2}$, so absorption controls
the shape of the X-ray spectrum up to $\sim 6$~keV. 
As discussed in \S4.2.1 of F99, the absorption seen in X-rays is
substantially larger than expected from the optical reddening;
the X-ray absorbing material may be dust poor.
The small 0.3--1~keV residuals left by the double partial-covering model 
may be due to $\sim 5\times 10^{40}$~ergs~s$^{-1}$ emission from ionized 
gas in the nucleus or host galaxy.
These residuals are consistent with X-ray emission from the Narrow Line 
Region, similar to that observed in NGC 4151 (Ogle et al.\ 2000) and the 
Seyfert 2 galaxies NGC 1068 (Ogle et al.\ 2003) and Mrk 3 (Sako et al.\ 2000). 
The residuals around 0.8--0.9~keV could be Fe~L emission, but this spectral 
complexity is difficult to model due to the limited low-energy photon statistics
of the EPIC instruments and the insufficient signal-to-noise ratio of the RGS spectra.

\begin{figure}[h!]
\centerline{ {\hfil\hfil
\psfig{figure=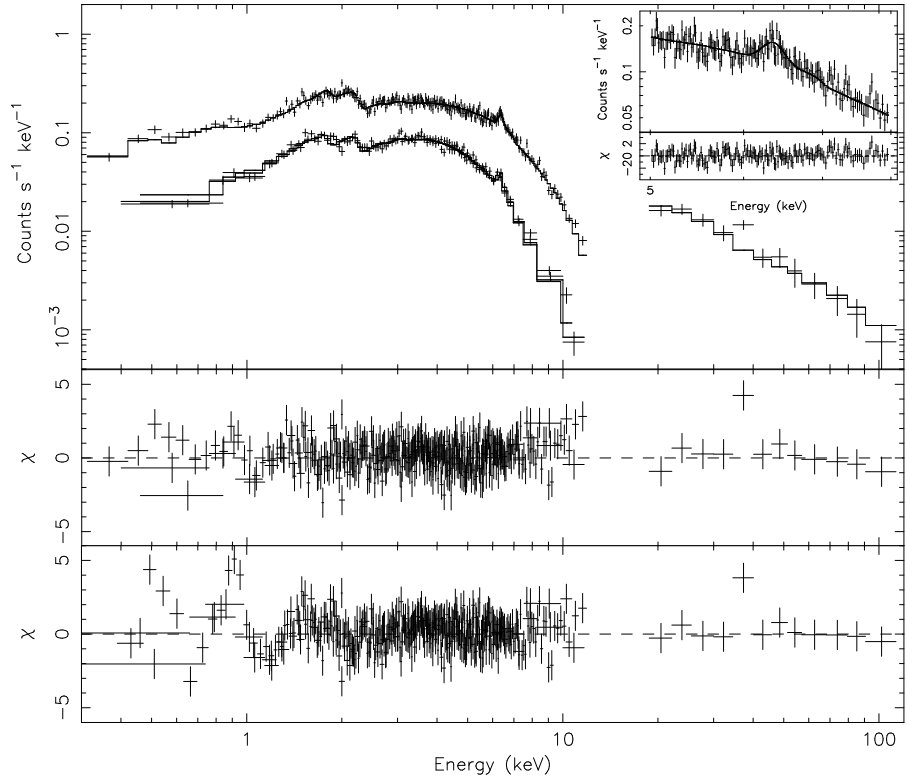,width=12cm,clip=,angle=180}
\hfil\hfil} }
\caption{\X\ EPIC p-n (top), MOS (lower) and \B\ PDS (right; $>18$~keV) spectra of 
Mrk~6 in the 0.3--120~keV band. The EPIC counts were binned into energy channels with 
a signal-to-noise ratio of ${\rm S/N}>10$ for plotting purposes. The best-fit double 
partial-covering model is plotted as the histograms (see model~2 in Tab.~\ref{spec_tab}).
The fit residuals from this model are shown in the middle panel (in units of $\sigma$).
The fit residuals from the partially ionized and neutral absorption model are given in 
the lower panel (see model~4 in Tab.~\ref{spec_tab}). The positive residual at
$\sim38$~keV appears to be statistical; we do not find any evidence for a calibration
error at this energy, and our results are not materially affected by the exclusion of
this data point.
The inset shows the (un-binned) 5--8~keV EPIC p-n spectrum around the Fe~K$\alpha$ line.
\label{spectrum}}
\end{figure}

Motivated by recent studies of NGC~4151 (Schurch \& Warwick 2002), we 
also tried fitting the measured X-ray absorption with another physically 
plausible model consisting of both ionized and neutral columns of gas. 
This model provides an acceptable fit to the X-ray spectrum above 2~keV, 
but it leaves substantially larger residuals at lower energies than does 
the double partial-covering model (visible in the lower panel of 
Fig.~\ref{spectrum}). Again these residuals may be plausibly explained
by emission from ionized gas in the nucleus or host galaxy.

Three pieces of evidence suggest that, at least at high energies, 
our X-ray observations have penetrated all the way to the black hole 
region of Mrk~6: 
(1) the 18--120~keV continuum shape measured by the \B\ PDS is 
consistent with those of Seyfert~1 galaxies (compare with
\S3.2 of Malaguti et~al. 1999),
(2) the relatively small equivalent width of the Fe~K$\alpha$ line
($EW=93^{+26}_{-20}$~eV) indicates significant dilution of the 
reflection continuum at 6.4~keV by direct power-law emission from 
the black hole region, and
(3) the relative 2--10~keV, [O~{\sc iii}], and far-infrared 
luminosities are consistent with those of Seyfert~1 galaxies
(see \S4.1 of F99). 
Given these results, we can have confidence that the 
2--10~keV 
($L_{2-10}=1.4\times10^{43}~{\rm ergs~s}^{-1}$ from \X) and 
20--100~keV 
($L_{20-100}=3.5\times10^{43}~{\rm ergs~s}^{-1}$ from \B) 
luminosities we have found for Mrk~6 represent those of the
intrinsic X-ray continuum (rather than just scattered X-ray
emission, for example). The X-ray luminosity of Mrk~6 is 
$\sim 5$ times the average X-ray luminosity of NGC~4151
($L_{2-10}\sim 3\times10^{42}~{\rm ergs~s}^{-1}$;
$L_{20-100}\sim 1\times10^{43}~{\rm ergs~s}^{-1}$;
Yang et~al. 2001; Schurch \& Warwick 2002; Schurch 2002).
Its lower X-ray flux arises only because it is $\sim 5.8$
times more distant. 

Our analyses have also revealed significant X-ray absorption variability
during the 1.5~yr between the \B\ and \X\ observations. In the
double partial-covering model, the column densities of both partial
covering components drop by a factor of $\sim2$ while the covering
fractions do not change significantly. If the fitted partial 
covering in fact indicates that we have both direct, absorbed
as well as electron-scattered, unabsorbed views of the nucleus, 
the observed changes seem physically plausible. The column densities 
along the direct line of sight could change due to gas motions in the 
torus atmosphere, while the scattered fraction would not change if 
scattering occurs on large spatial scales. The mechanism of absorption 
variability must allow the column density to undergo large fractional 
changes; such changes could perhaps arise due to bulk rotation of a 
torus or Poisson fluctuations in a relatively small number of obscuring 
`clouds' along the line of sight. Our discovery of X-ray absorption 
variability from Mrk~6 (as well as the observed variability amplitude 
and timescale) is generally consistent with the absorption variability 
seen from other absorbed Seyferts (e.g., Risaliti, Elvis, \& Nicastro 2002; 
Schurch \& Warwick 2002). 

Based on our results, further monitoring of the X-ray absorption 
variability of Mrk~6 is merited to refine understanding of its
amplitude and timescale. Such monitoring could be significantly
complemented by coordinated optical spectroscopy, given the known 
optical line profile variability (see \S1). High-quality ultraviolet 
spectroscopy of Mrk~6 would allow connections to be made between the
X-ray absorption and any ultraviolet absorption. The archival
{\sl IUE} data on Mrk~6 suggest intrinsic Mg~{\sc ii} absorption, 
but higher quality data are required for a proper study.  

\acknowledgments

We thank L.A. Antonelli,  M.C. Eracleous, and N.J. Schurch for helpful discussions.
The project was supported by NASA grants NAG5-9940 and LTSA NAG5-8107 and
NAG5-13035.



\clearpage

\begin{deluxetable}{lcccccc}
\tablenum{1}
\tabletypesize{\footnotesize}
\tablecaption{X-Ray Observations of Markarian~6 \label{obs_tab}}
\tablewidth{0pt}
\tablehead{
\colhead{} & 
\colhead{} & 
\colhead{Observation} & &
\colhead{Exposure} \\
\colhead{Observatory} & 
\colhead{Instr.} & 
\colhead{ID} &
\colhead{Date$^{\rm a}$} & 
\colhead{(ks)}}
\startdata
\A\ & GIS 	& 75041000 & 1997-04-07 & \phantom{0}42.1 \\
\A\ & SIS 	& 75041000 & 1997-04-07 &  \phantom{0}36.4 \\\noalign{\medskip}
\B\ & LECS	& 51067001 & 1999-09-14 & \phantom{0}49.3 \\
\B\ & MECS	& 51067001 & 1999-09-14 & 109.4 \\
\B\ & PDS	& 51067001 & 1999-09-14 & \phantom{0}52.0 \\
\noalign{\medskip}
\X\ & p-n	& 0061540101 & 2001-03-27 &  \phantom{0}31.8 \\
\X\ & MOS~1	& 0061540101 & 2001-03-27 &  \phantom{0}30.8 \\
\X\ & MOS~2	& 0061540101 & 2001-03-27 &  \phantom{0}30.8 \\
\X\ & RGS~1	& 0061540101 & 2001-03-27 &  \phantom{0}46.4 \\
\X\ & RGS~2	& 0061540101 & 2001-03-27 &  \phantom{0}37.9 \\
\noalign{\smallskip}
\enddata
\tablenotetext{(a)}{\phantom{0}Start of observation.}
\end{deluxetable}

\vfill


\begin{deluxetable}{lcccccc}
\tablenum{2}
\tabletypesize{\footnotesize}
\tablecaption{Spectral Fitting Parameters of Markarian~6 \label{spec_tab}}
\tablewidth{0pt}
\tablehead{
\noalign{\medskip}
\colhead{} &
\multicolumn{4}{c}{Model Description and Data Used$^{\rm a}$} \\
\colhead{} &
\colhead{(1)} &
\colhead{(2)} &
\colhead{(3)} &
\colhead{(4)} \\
\colhead{} & \cline{1-4} \\
\colhead{Parameter} &
\colhead{PL} &
\colhead{PL+2PC+GA} &
\colhead{RE+2PC+GA} &
\colhead{RE+PI+GA} & \\
&
\colhead{(PDS)} &
\colhead{(PDS+p-n+MOS)} &
\colhead{(PDS+p-n+MOS)} &
\colhead{(PDS+p-n+MOS)}
 }
\startdata
Absorption, $N_{\rm H}$ ($10^{22}~{\rm cm}^{-2}$)\dotfill 
	& $\cdots$
	& $0.11^{+0.01}_{-0.01}$
	& $0.11^{+0.01}_{-0.01}$
	& $0.34^{+0.02}_{-0.01}$ \\
\noalign{\medskip}
Photon index, $\Gamma$\dotfill 
	& $1.81^{+0.22}_{-0.20}$
	& $1.81$ (fixed)
	& $\cdots$
	& $\cdots$ \\
$A1$ (${\rm cts~keV}^{-1}~{\rm cm}^{-2}~{\rm s}^{-1}$)\dotfill 
	& $9.48^{+0.12}_{-0.10} \times 10^{-2}$ 
	& $5.39^{+0.05}_{-0.06} \times 10^{-3}$ 
	& $\cdots$ 
	& $\cdots$ \\
\noalign{\medskip}
Partial covering, $N_{\rm H}^1$ ($10^{22}~{\rm cm}^{-2}$)\dotfill 
	& $\cdots$
	& $10.96^{+0.58}_{-0.42}$
	& $8.05^{+0.69}_{-0.48}$
	& $\cdots$ \\
Partial covering, $f_{\rm c}^1$ ($\%$)\dotfill 
	& $\cdots$
	& $57^{+1}_{-1}$
	& $56^{+2}_{-1}$
	& $\cdots$  \\
Partial covering, $N_{\rm H}^2$ ($10^{22}~{\rm cm}^{-2}$)\dotfill 
	& $\cdots$
	& $2.46^{+0.07}_{-0.05}$
	& $2.27^{+0.03}_{-0.08}$
	& $\cdots$ \\
Partial covering, $f_{\rm c}^2$ ($\%$)\dotfill 
	& $\cdots$
	& $93^{+1}_{-1}$
	& $91^{+1}_{-1}$
	& $\cdots$ \\
\noalign{\medskip}
Photon index, $\Gamma$\dotfill 
	& $\cdots$ 
	& $\cdots$ 
	& 1.81 (fixed)
	& 1.81 (fixed) \\
Cutoff energy (keV)\dotfill 
	& $\cdots$ 
	& $\cdots$ 
	& $238^{+499}_{-54}$
	& $2735^{+383}_{-2368}$ \\
Inclination angle, cos~$i$ \dotfill 
	& $\cdots$ 
	& $\cdots$ 
	& $0.94^{+0.01}_{-0.25}$
	& $0.95^{+0.01}_{-0.19}$ \\
Reflection fraction, $r$ \dotfill 
	& $\cdots$
	& $\cdots$ 
	& $1.21^{+0.14}_{-0.16}$
	& $1.33^{+0.21}_{-0.01}$ \\
$A2$ (${\rm cts~keV}^{-1}~{\rm cm}^{-2}~{\rm s}^{-1}$)\dotfill 
	& $\cdots$ 
	& $\cdots$ 
	& $4.68^{+0.02}_{-0.02} \times 10^{-3}$
	& $4.59^{+0.03}_{-0.06} \times 10^{-3}$ \\
\noalign{\medskip}
Warm absorption, $N_{\rm H}$ ($10^{22}~{\rm cm}^{-2}$)\dotfill 
	& $\cdots$ 
	& $\cdots$
	& $\cdots$
	& $13.32^{+0.12}_{-0.13}$ \\
Ionization parameter, ${\rm log}~\xi$ \dotfill
	& $\cdots$
	& $\cdots$ 
	& $\cdots$ 
	& $2.42^{+0.01}_{-0.01}$ \\
\noalign{\medskip}
Line energy, $E$ (keV)\dotfill
	& $\cdots$
	& $6.45^{+0.03}_{-0.04}$
	& $6.47^{+0.03}_{-0.03}$
	& $6.45^{+0.04}_{-0.03}$  \\
Line $EW$ (eV)\dotfill  
	& $\cdots$
	& $93^{+26}_{-20}$
	& $87^{+20}_{-26}$
	& $84^{+19}_{-26}$ \\
$A3$ (${\rm cts~cm}^{-2}~{\rm s}^{-1}$)\dotfill 
	& $\cdots$ 
	& $1.85^{+0.60}_{-0.37} \times 10^{-5}$
	& $1.68^{+0.39}_{-0.53} \times 10^{-5}$
	& $1.67^{+0.41}_{-0.33} \times 10^{-5}$ \\
\noalign{\medskip}
$\chi^2/{\rm d.o.f.}~(\chi^2_{\nu})$ \dotfill 
	& 3.61/9 (0.40) 
	& 824.2/858 (0.96)
	& 812.8/855 (0.95)
	& 937.3/863 (1.09) \\
\enddata
\tablecomments{All model fits include Galactic absorption by a column density
of $N_{\rm H}=6.4\times10^{20}~{\rm cm}^{-2}$.}
\tablenotetext{(a)}{PL: power law; PC: partial covering; GA: Gaussian emission line;
	  PI: photoionized absorption; RE: power law with Compton reflection 
          by neutral material (PEXRAV).}
\end{deluxetable}



\clearpage

\end{document}